\documentstyle{article}
\font\msbm=msbm10 scaled \magstep1
\newcommand{\Bbb}[1]{\mbox{\msbm #1}\,}
\def\be{\begin{equation}}
\def\ee{\end{equation}}
\def\bean{\begin{eqnarray}}
\def\eean{\end{eqnarray}}
\def\bea*{\begin{eqnarray*}}
\def\eea*{\end{eqnarray*}}
\def\ve{\varepsilon}
\def\pt{\partial}
\def\ba{\begin{array}}
\def\ea{\end{array}}
\def\vf{\varphi}
\def\t{\theta}
\def\nn{\nonumber}
\def\s{\stackrel}
\def\o{\omega}
\def\si{\sigma}

\begin{document}
\begin{center}
{\Large\bf On the asymptotic  reduction to the multidimensional
nonlinear Schrodinger equation}
\footnote {This work was partially supported by
Russian Fund of Basic Research (grants 96-15-96241, 97-01-00459)}
\vspace*{3mm}\\
M.M. Shakir'yanov\\
\vspace*{3mm}
Institute of Mathematics of the Russian Academy of Sciences \\
112 Chernyshevskii str., Ufa 450000, RUSSIA \\ e-mail:
marsh@imat.rb.ru
\end{center}


\begin {abstract}
\noindent
The problem on the asymptotics for the solution of
multidimensional nonlinear Boussinesq equation with respect to a
small parameter $\ve$ is considered. The asymptotic expansion of
the solution  of this problem with respect to $\ve\to0$ for long
times $t\sim {\cal O}(\ve^{-2})$ is constructed and justified.
The leading terms of the asymptotic solution are defined from the
multidimensional nonlinear Schrodinger equation and from the
linear homogeneous wave equation.
\end {abstract}

\section * {1 Introduction}

In this work the problem on the asymptotic reduction in
multidimensional waves is investigated where the construction of
the asymptotic solution reduces to the multidimensional nonlinear
Schrodinger equation.

The problem on the asymptotics with respect to the small
parameter $\ve$ for the multidimensional nonlinear Boussinesq
equation is considered:
\be
\pt_{t}^{2}u-\Delta_{x{\bar y}}u+\nu [\pt_{x}^{4}+ \pt_{{\bar y}}^{4}] u+
\hbox{div}\ {\bf g}(\nabla u)=0,\quad x\in{\Bbb R},\
{\bar y}\in{\Bbb R}^{n-1},\  t>0,
\label{eq3.10}
\ee
Here
$
\nabla=(\pt_{x},\nabla_{{\bar y}}),\
\nabla_{{\bar y}}=(\pt_{y_{2}},\dots,\pt_{y_{n}}),\
\Delta_{{\bar y}}=\nabla_{{\bar y}}^{2},\
\Delta_{x{\bar y}}=\pt_{x}^{2}+\Delta_{{\bar y}},
$
$
\pt_{{\bar y}}^{4}=\pt_{y_{2}}^{4}+\dots+\pt_{y_{n}}^{4},\
\hbox{div}\ {\bf g}=<\nabla\cdot {\bf g}>,\
\nu=\hbox{const}>0.
$
The initial conditions are taken as a combination of plane waves
with small amplitudes:
\be
\left[ \ba{c} u \\ u_{t} \ea \right]
(x,{\bar y},t,\ve)_{\mid {t=0}}=
\ve\sum \limits _{k=0,\pm 1}
\Lambda _{k}e^{ikx},
\label{eq3.11}
\ee
and deformed by slow variables:
$$
\Lambda _{k}=
\left[ \ba{c} \vf_{k} \\ \psi _{k} \ea \right](\ve x,\ve {\bar y}),
\quad \Lambda^{*}_{k}=\Lambda_{-k},
\quad \psi _{0}\equiv 0.
$$
Their components are considered as functions which decrease fast
at infinity:
\be
\vf_{k}, \psi _{k}(\xi,{\bar \eta})=
{\cal O}((|\xi |+|{\bar \eta}|)^{-N}),\quad |\xi|+|{\bar
\eta}|\to\infty,
\quad \forall N>0, \quad \forall k.
\label{vanish}
\ee
The purpose of this work is the construction and justification of
asymptotic expansion of the solution $u (x,{\bar y},t,\ve)$ of
the Cauchy problem (\ref {eq3.10})--(\ref {eq3.11}) as $ \ve \to
0$, uniformly on a long-time interval $ 0\leq t\leq {\cal O}
(\ve^ {-2}), \ \forall (x, {\bar y}) \in {\Bbb R}^{n}.$

The main result is the following. The determination of leading
terms of the asymptotic solution of the problem (\ref
{eq3.10})--(\ref {eq3.11}) is reduced to the solution of the
Cauchy problems for the multidimensional nonlinear Schrodinger
equation (NLS):
\be
i\pt _{\t}w+ \alpha
\pt _{\si}^{2}w+\delta
\Delta _{{\bar \eta}}w
+\gamma w|w|^{2}=0,
\quad \t>0,
\label{eqnls}
\ee
\be
w(\si , {\bar \eta},\t)_{|\t =0}=w_{0}(\si ,{\bar \eta}),
\quad (\si,{\bar \eta})\in{\Bbb R}^{n},
\label{eqind}
\ee
and for the linear homogeneous wave equation:
\be
\pt _{\tau}^{2}\phi-\Delta_{\xi{\bar \eta}}
\phi=0,\ \tau>0; \quad
\phi_{|\tau =0}=
\vf_{0}(\xi ,{\bar \eta}),\quad
\pt _{\tau} \phi_{|\tau =0}=0,
\quad (\xi ,{\bar \eta} )\in{\Bbb R}^{n}.
\label{wave}
\ee

The formal constructions of the asymptotic solution of the Cauchy
problem (\ref {eq3.10})--(\ref {eq3.11}) are obtained here
without any essential restrictions of the input data. We only
suppose that the components of vector function $ {\bf g}$ are
decomposed into the asymptotic Taylor serieses as $ | v | + |
\bar {w} | \to 0\ ( v\in {\Bbb R}, \ {\bar w} \in {\Bbb R} ^
{n-1}) $ with specific quadratic terms: \footnote {Here summation
is conducted along coinciding indexes}
\be
g_{j}(v,\bar {w})=g_{j,i}^{0,2}w_{i}^{2}+g_{j}^{3,0}v^{3}+
g_{j,i}^{2,1}v^{2}w_{i}+ g_{j,i}^{1,2}vw_{i}^{2}+
g_{j,i}^{0,3}w_{i}^{3}+\dots,\ j=\overline{1,n}.
\label{eqt}
\ee

It is worth noting that in case of general nonlinearity the
formal constructions reduce to the Davey-Stewartson type systems
of equations \cite {lk3}.

Justification of the asymptotic expansion will be carried out
under the side conditions which are connected with the proof of
solvability of the input problem. One of the requirements is the
absence of the part of the cubic terms in expansion (\ref {eqt}):
\be
g_{j}^{3,0}\equiv0,\quad j=\overline{2,n}.
\label{eqt1}
\ee

For the two-dimensional case $(n=2)$ the formal asymptotic
reductions to the NLS and Davey-Stewartson type systems of
equations are known from \cite {ds2, djred1, fred1}.
Justification of these asymptoticses for multidimensional
problems was obtained only by few authors. For example,
justification of reduction to the "shallow water" equations for
2+1-dimensional surface waves in a class of analytical functions
is obtained in \cite {ovs1}; justification of reduction to the
Kadomtsev-Petviashvili equation is reduced in \cite {lk3}.

In the present work we justify the asymptotic reduction to the
multidimensional NLS. The main mathematical result consists of
solvability of the input problem and evaluations of the residual
of the asymptotic expansion. The obtained result gives the exact
sense of asymptotic reduction to the multidimensional NLS. It is
worth mentioning that here we use the ideas and sentences
expressed in \cite {lk3, lk1}.

The problem (\ref {eq3.10})--(\ref {eq3.11})  is considered as a
simplest example where the construction of asymptotic solution
reduces to the multidimensional NLS. Similar results may be
obtained and for the more complicated equations of type (\ref
{eq3.10}), generally, for pseudodifferential ones.

Here we shall be limited to reviewing of the equation (\ref
{eq3.10}) for revealing basic singularities of the construction
and justification of the asymptotic expansion in multidimensional
problems.

\newtheorem {theor} {Theorem}
\begin {theor} \hskip-2mm {\bf}.
Let's suppose the functions $ g _ {j} (\nabla u), \ j = \overline {1, n} $ are
analytical in neighbourhood of zero so Taylor serieses  (\ref {eqt}), (\ref
{eqt1}) converge at $ | \nabla u | \leq M $. Initial functions $\vf _ {k},
\psi _ {k} (\xi, {\bar \eta}), \ k = 0, \pm 1 $ satisfy to (\ref
{vanish}). Let them be analytical in layer $S(\beta _ {0})=
\left\{\vert Im\xi\vert, \vert Im\eta _ {2} \vert, \dots, \vert
Im\eta _ {n} \vert
\leq\beta _ {0} \right\} $ $ \quad (\beta _ {0} = const > 0) $. Then there are
the values $ \ve _ {0}, T> 0, \ \beta \in ( 0, \beta _ {0}):\
\forall\ve
\in (0, \ve _ {0}) $ the Cauchy problem (\ref {eq3.10})--(\ref {eq3.11}) has a
unique solution $ u (x, {\bar y}, t, \ve)$ in $ \Omega (T) =\left\{ (x,{\bar
y}) \in {\Bbb R} ^ {n}, 0\leq t\leq T\ve ^ {-2} \right \} $ which is
analytical in layer $ S (\beta)$. This solution has the following asymptotics:
$$
u = \ve \Big [v _ {0} (\xi, {\bar \eta}, \tau) +
\sum\limits _ {k = \pm1} \sum\limits _ {\o = \pm\o (k)}
w_{k, \o} (\si, {\bar \eta}, \t) e ^ {ikx-i\o t} \Big] + {\cal O}
(\ve^ {2}),
$$
$$
\xi = \ve x, \ {\bar \eta} = \ve {\bar y}, \ \tau = \ve t, \
\t = \ve ^ {2} t, \ \si = \xi-\o ^ {'} \tau, \ \o (k) = k\sqrt {1+\nu k^{2}},
$$
uniformly in $\Omega (T)$. The complex amplitudes $ w = w _ {k,
\o} $ are the solutions of the Cauchy problem (\ref{eqnls})--(\ref
{eqind}) for the nonlinear Schrodinger equations with constants:
$ \alpha = \o ^ {"} /2, \
\delta = 1 / (2\o), \ \gamma = 3g _ {1} ^ {3,0} / (2\o) $ and with initial
value: $w_{0} = 1/2\vf _ {k} + i\delta \psi _ {k}. $ The
amplitude of zero harmonic $\phi = v _ {0}$  is defined from the
Cauchy problem (\ref {wave}) for the homogeneous wave equation.
\label {th0}
\end {theor}
The functions $ w _ {k, \o} (\si, {\bar \eta}, \t)$ represent
slowly deformed (in a scale $ \t = \ve ^ {2} t $) amplitudes of
wave packets travelling with group velocity $\o ^ {'} $ in
characteristic directions $ x-\o
^ {'} t = \hbox {const}.$ These amplitudes are modulated both in longitudinal
($ \si = \ve (x-\o ^ {'} t) $) and in cross $ ({\bar \eta} = \ve {\bar y}) $
directions.

\section*{2 The formal constructions}

For the formal construction of the asymptotic solution of the
problem (\ref{eq3.10})--(\ref {eq3.11}) we shall use the method
of multiply scale \cite {naif1}. For this purpose we use the
above entered slow variables $ \xi, {\bar \eta}, \tau, \t.$ We
shall search for a solution in the form of $u (x, {\bar
y},t,\ve)=v (x,\xi,{\bar
\eta}, t, \tau, \t, \ve). $ Taking into account the fact that
derivatives will be transformed by the rules:
\bea*
\pt_{t}u &= &\pt _{t}v+\ve \pt _{\tau}v+\ve ^{2}\pt _{\t}v
\equiv D_{t}v,\\
\pt_{x}u &=& \pt _{x}v+\ve \pt _{\xi}v\equiv D_{x}v, \\
\nabla _{{\bar y}}u &=&\ve\nabla _{{\bar \eta}}v,
\eea*
we shall obtain the following equation for $v$:
\bean
[\pt _{t}^{2}-\pt _{x}^{2}+\nu \pt _{x}^{4}]v &+& 2\ve [\pt
_{t}\pt _{\tau}-\pt _{x}\pt _{\xi}+ 2\nu \pt _{x}^{3}\pt
_{\xi}]v + \nn \\ &+& \ve ^{2}[\pt _{\tau}^{2}+2\pt _{t}\pt
_{\t}-\pt _{\xi}^{2}-\Delta _{{\bar \eta}}+
6\nu\pt _{x}^{2}\pt _{\xi}^{2}]v+ \nn \\
&+& 2\ve ^{3}[\pt _{\tau}\pt _{\t} +
2\nu \pt _{x}\pt _{\xi}^{3}]v+
\ve ^{4}[\pt _{\t}^{2}+\nu \pt _{\xi}^{4}+
\nu  \pt_{{\bar \eta}}^{4}]v+ \nn \\
&+& D_{x}g_{1}(D_{x}v,\ve
\nabla_{{\bar \eta}}v)+\ve<\nabla_{{\bar \eta}}\cdot
\bar {g}(D_{x}v,\ve \nabla_{{\bar \eta}}v)>=0
\label{eq3.12}
\eean
with the initial conditions:
\be
\left[ \ba{c} v \\ D_{t}v
\ea \right]
_{\mid {t=\tau=\t=0}}=
\ve\sum \limits _{k=0,\pm 1}
\Lambda _{k} (\xi,{\bar \eta})e^{ikx}.
\label{eq3.13}
\ee

The {\it formal asymptotic solution} \ (FAS) of the Cauchy
problem (\ref {eq3.12})--(\ref {eq3.13}) is called  a segment of
series
\be
v\sim \ve\sum \limits _ {n = 0} ^ {\infty} \ve ^ {n}
\s {n} {v} (x, t, \xi, {\bar \eta}, \tau, \t),
\label {eq3.14}
\ee
which, first of all, satisfies to equation (\ref {eq3.12}) with
exactitude $ {\cal O}(\ve^{4}) $ and to the initial values (\ref
{eq3.13}) with exactitude $ {\cal O}(\ve^{3}) $ and, secondly,
each consequent term of this segment is less than the previous
one in an $\ve$ order, uniformly in the corresponding domain with
respect to independent variables.

By the method explained below we may construct any length of the
series's segment (\ref {eq3.14}), sequentially defining
dependence of coefficients of asymptotics in all temporal scales.
Only three terms of this series are constructed here. It is
enough for deriving the required residuals. The leading term $\s
{0} {v}$ of the asymptotics is finally defined dependending on
variables $t,\tau,\t;$ the first correction $\s {1} {v} $ is
completely defined in scales $t,\tau,$ the second correction $\s
{2} {v} $
- in scale $t.$

The coefficients $\s {n}{v}$ are constructed in the form of final
Fourier sums:
\be
\s {n}{v}= \sum \limits _{k,\o}\s {n}{v}_{k,\o}
(\xi, {\bar \eta}, \tau, \t)e^{ikx-i\o t},
\quad (\s {n}{v}_{k,\o}=\s {n}{{v}^{*}}_{-k,-\o}),
\label{eq3.15}
\ee
where sign $*$ denotes complex conjugate. The problem is reduced
to defining the coefficients $\s{n}{v}_ {k,\o} $.

Substituting serieses (\ref {eq3.14})--(\ref {eq3.15}) into the
equation (\ref {eq3.12}) and equating terms at identical degrees
of parameter $ \ve $ and identical indexes $k,\o,$ we shall
obtain the equations on $ \s {n} {v} _ {k,
\o} $. So, on the first step, at $\ve^ {1} $ we obtain the homogeneous
equations on $\s {0}{v}$:
$$
[ \pt _ {t} ^ {2} -\pt _ {x} ^ {2} + \nu \pt _ {x} ^ {4}] \s {0} {v} = 0.
$$
Dispersion relation is defined from here:
$$
\o ^ {2} -k ^ {2} -\nu k ^ {4} = 0,
$$
whence $\o=\pm \o (k) \equiv\pm k\sqrt {1 + \nu k ^ {2}}.$ The
values $ k = 0, \pm 1 $ are taken according to the initial data
(\ref {eq3.13}).
\par
\noindent
Hence, we have the following representation for $\s {0}{v}$:
$$
\s {0} {v} = v _ {0} (\xi, {\bar \eta}, \tau, \t) +
\sum \limits _ {k = \pm 1} \sum \limits _ {\o = \pm \o (k)}
\s {0} {v} _ {k, \o} (\xi, {\bar \eta}, \tau, \t)
e^{ikx-i\o t}.
$$
The initial conditions (\ref {eq3.13}) give values of the
functions $v_{0} $ and $\s {0} {v} _ {k, \o} $ at $ \tau = \t =
0:$
$$
v_{0\mid \tau = \t = 0} = \vf _ {0} (\xi, {\bar \eta}),
$$
\be
\ba{ll}
&[ \s {0}{v}_{k,\o (k)}+
\s {0}{v}_{k,-\o (k)}]_{\mid \tau=\t =0}=
\vf _{k}(\xi,{\bar \eta}), \nn \\
-i\o (k) &[ \s {0}{v}_{k,\o (k)}-
\s {0}{v}_{k,-\o (k)}]_{\mid \tau=\t =0}=
\psi _{k}(\xi,{\bar \eta}), \quad k=\pm 1.
\ea
\label{eq3.17}
\ee

On the second step, equating expressions at $\ve^{2}$, we shall obtain the
inhomogeneous equation:
\be
- [\pt _ {t} ^ {2} -\pt _ {x} ^ {2} + \nu \pt _ {x} ^ {4}] \s {1} {v} =
2 [\pt _ {t} \pt _ {\tau} -\pt _ {x} \pt _ {\xi} + 2\nu \pt _ {x} ^ {3} \pt
_ {\xi}] \s {0} {v}.
\label {eq3.18}
\ee
In order that solution $ \s {1} {v} $ shouldn't have secular
(growing at $ t\to\infty $) terms it is necessary to equate to
zero the functions in right-hand side of (\ref {eq3.18}) which
are the solutions of the corresponding homogeneous equation.
Whereas
$$
[\pt _{t}\pt _{\tau}-\pt _{x}\pt _{\xi}+ 2\nu \pt _{x}^{3}\pt
_{\xi}]\s{0}{v}=
-\sum \limits _{k=\pm 1}\sum \limits _{\o=\pm \o (k)}
i\o e^{ikx-i\o t} [\pt _{\tau}+\o ^{'}(k)\pt _{\xi}]\s
{0}{v}_{k,\o},
$$
the right-hand side of (\ref {eq3.18}) consists completely of
solutions of the homogeneous equation. Therefore they should be
excluded from the requirements:
\be
[ \pt _ {\tau} + \o ^ {'} (k) \pt _ {\xi}] \s {0} {v} _ {k, \o} = 0,
\quad k = 1, \ \o = \pm \o (k), \quad \xi\in {\Bbb R}, \ \tau > 0.
\label {eq3.181}
\ee
These equations are trivial and allow in functions $ \s {0} {v} _ {k, \o}, \
\forall k\neq 0 $ to define a structure of dependence from $ \xi, \tau $ (on
the first slow scale) for all values $\tau\geq0:$
$$
\s {0} {v} _ {k, \o} (\xi, {\bar \eta}, \tau, \t) =
w _ {k, \o} (\si _ {k, \o}, {\bar \eta}, \t), \quad
\si _ {k, \o} = \xi -\o ^ {'} \tau,
\quad k = 1, \ \o = \pm \o (k).
$$
Thus the number of independent variables is diminished in the
coefficients of the leading term of the asymptotics. The initial
conditions (\ref {eq3.17}) define functions $w_{k,\o}$ at $\t =
0:$
$$
w_{k,\o}(\si _{k,\o}, {\bar \eta},\t)_{|\t =0}=
\frac{1}{2} \left( \vf _{k}+
\frac{i}{\o} \psi _{k}\right)(\si _{k,\o},{\bar \eta}),\quad
k=1,\ \o=\pm \o (k).
$$
The values of these functions as well as the leading term of the
asymptotics remain undefined from slow dependence on time $\t$
(in the second scale). In order to define dependence on $\t$ it
is necessary to analyze the next corrections of the asymptotic
solution.

After the carried out elemination of the secular terms first the
correction satisfies to the equation:
\be
[ \pt _ {t} ^ {2} -\pt _ {x} ^ {2} + \nu \pt _ {x} ^ {4}] \s {1} {v} = 0.
\label {eq3.20}
\ee
The particular solution of the equation (\ref {eq3.20}) is trivial: $\s {1}
{v} = 0.$

The general solution of the equation (\ref {eq3.20}) is
supplemented by the solution of the homogeneous one:
$$
\s {1} {v} = \sum \limits _ {k = \pm 1, \pm 2} \sum
\limits _ {\o = \pm \o (k)} \s {1} {v} _ {k, \o} (\xi, {\bar \eta}, \tau, \t)
e ^ {ikx-i\o t}
$$
with undefined coefficients $\s {1}{v}_{k, \o}$ for the time
being.

The initial conditions (\ref {eq3.13}) in an $ \ve ^ {2}$ order
give initial values for these functions and also for
$\pt_{\tau}v_{0}$:

for $ k = 0 $:
$$
\pt _ {\tau} v _ {0} (\xi, {\bar \eta}, \tau, \t)
_ {\mid \tau = \t = 0} = 0;
$$

for $ k = \pm1 $:
\be
\ba{ll}
&[\s {1}{v}_{k,\o (k)}+
\s {1}{v}_{k,-\o (k)}]_{\mid \tau=\t =0}=0, \nn \\
-i\o (k) &[ \s {1}{v}_{k,\o (k)}-
\s {1}{v}_{k,-\o (k)}]_{\mid \tau=\t =0}=
-\pt _{\tau}
[\s {0}{v}_{k,\o (k)}+
\s {0}{v}_{k,-\o (k)}]_{\mid \tau=\t =0};
\ea
\label{eq3.21}
\ee

for $ k = \pm2 $:
\be
\ba{ll}
&[\s {1}{v}_{k,\o (k)}+
\s {1}{v}_{k,-\o (k)}]_{\mid \tau=\t =0}=0, \nn \\
-i\o (k) &[ \s {1}{v}_{k,\o (k)}-
\s {1}{v}_{k,-\o (k)}]_{\mid \tau=\t =0}=0.
\ea
\label{eq3.22}
\ee

The amplitudes $ \s {1} {v} _ {k, \o (k)}, \ k=1,2, \  \o = \pm\o
(k), $ on this step remain undefined in dependence on $ \tau, \
\t > 0$. The amplitude of  zero harmonic $v_{0}$  remains undefined
too. The equations for them will be obtained on the following
step.

In an $\ve^{3}$ order we obtain a linear inhomogeneous equation:
\bean
-[\pt _{t}^{2}-\pt _{x}^{2}+\nu \pt _{x}^{4}]\s{2}{v}&=&
2 [\pt _{t}\pt _{\tau}-\pt _{x}\pt _{\xi}+ 2\nu \pt _{x}^{3}\pt
_{\xi}]\s{1}{v}+ \nn \\ &+& [\pt _{\tau}^{2}-\pt _{\xi}^{2}-\Delta
_{{\bar \eta}}+ 2\pt _{t}\pt _{\t}+6\nu \pt _{x}^{2}\pt
_{\xi}^{2}]
\s{0}{v}+ \nn \\
&+& g_{1}^{3,0} \pt _{x}(\pt _{x}\s{0}{v})^{3},
\label{eq3.23}
\eean
from the right-hand side of which it is necessary to eliminate
secular addends. The elemination of secularities reduces to the
equations for amplitudes $ \s {1} {v} _ {k, \o}, \ k = 1, \ \o =
\pm\o (k)$ of the first  correction in scale $\xi, \tau:$
$$
2i\o [\pt
_ {\tau} + \o ^ {'} (k) \pt _ {\xi}] \s {1} {v} _ {k, \o} = [-2i\o \pt
_ {\t} + ((\o ^ {'}) ^ {2} -1-6\nu k ^ {2})
\pt _ {\xi} ^ {2} -\Delta _ {{\bar \eta}}] w _ {k, \o} -
$$
\be
-3g _ {1} ^ {3,0} w _ {k, \o} [| w _ {k, \o} | ^ {2} + 2 | w_{k,-\o} |^{2}],
\quad k = 1, \ \o = \pm\o (k).
\label {eq3.24}
\ee
On the same step we obtain the equations for the amplitude of
zero harmonic of the leading term:
\be
\pt _ {\tau} ^ {2} v _ {0} -\Delta _ {\xi {\bar \eta}} v _ {0} = 0.
\label {eq3.25}
\ee
We have obtained the initial conditions for $v_{0}$ on the
previous steps, namely:
\be
\left [
\ba {c}
v_{0} \nn \\ \pt _ {\tau} v _ {0}
\ea
\right] _ {| \tau = \t = 0} =
\left [
\ba {c}
\vf_{0}(\xi, {\bar \eta}) \\ 0
\ea
\right], \quad (\xi, {\bar \eta}) \in {\Bbb R} ^ {n}.
\label {eq3.26}
\ee

So, on the first slow scale $ \tau $ 0 the amplitudes of the
leading term of the asymptotics are defined from the equations
(\ref {eq3.181}), (\ref {eq3.25}). These equations are linear.
Just this circumstance allows "to attain" up to long times $
t\approx\ve ^ {-2} $ in formal solution. To guarantee the absence
of secularities in first two orders of the formal solution we
should analyze the equations (\ref {eq3.24}), (\ref {eq3.25})
their solutions' boundedness at $\tau\to\infty.$

It is known in case of rapidly decreasing at infinity of the
initial data (\ref {eq3.26}), the solution of the Cauchy problem
for two-dimensional $ (n = 2) $ wave equation (\ref {eq3.25})
decreases at $\tau\to\infty$ \cite {lksh3}. The result of the
uniformly boundedness of the solution of the Cauchy problem (\ref
{eq3.25})--(\ref {eq3.26}) for the case of multidimensions will
be represented in the following section.

Let's pass to the equations (\ref {eq3.24}). Analyzing their
solutions we obtain the equations for $w _ {k, \o} (\si_ {k,
\o}, {\bar\eta}, \t) \ (\si _ {k, \o}=
\xi-\o ^ {'} \tau, k = 1, \ \o = \pm\o (k)).$ The right-hand side of
the equations (\ref {eq3.24}) contains the solutions of
homogeneous one as functions depending on $ \si = \si _ {k,\o}
=\xi-\o ^ {'}\tau.$ In order that solutions of inhomogeneous
equations (\ref {eq3.24}) shouldn't contain secular terms we must
eliminate the above  mentioned functions. Equating to zero of
expressions, depending on $\si = \si_ {k, \o} $ reduces to
nonlinear Schrodinger equations $ (k = 1, \ \o = \pm\o (k)): $
\be
[ -2i\o \pt _ {\t} + ((\o ^ {'}) ^ {2} -1-6\nu k ^ {2})
\pt _ {\si} ^ {2} -\Delta _ {{\bar \eta}}] w _ {k, \o} =
3g _ {1} ^ {3,0} w _ {k, \o} | w _ {k, \o} | ^ {2}.
\label {eq3.30}
\ee

So, in characteristic directions $ \si = \si _ {k, \o} = \xi-\o ^
{'} \tau = \hbox {const} $ at $ | \xi | + \tau \to \infty $ the
amplitudes of the leading term of the formal solution are defined
from  (\ref {eq3.25}) and (\ref {eq3.30}). According to
$$
\frac{((\o ^{'})^{2}-1-6\nu k^{2})}{-\o}=
\frac{1+6\nu k^{2}}{\o}-\frac{(k+2\nu k^{3})^{2}}
{\o ^{3}}=\frac{d^{2}\o}{dk^{2}} \equiv \o ^{''},
$$
the equation (\ref {eq3.30}) for $w=w_{k,\o}$ has the form:
\be
i\pt _ {\t} w + \alpha
\pt _ {\si} ^ {2} w + \delta
\Delta _ {{\bar \eta}} w
+ \gamma w | w | ^ {2} = 0,
\label {eq3.35}
\ee
where $ \alpha = \o ^ {"} /2, \ \delta = 1 / (2\o), \ \gamma = 3g
_ {1} ^ {3,0} / (2\o). $  The equation (\ref {eq3.35}) describes
the slow (in scale $ \t = \ve ^ {2} t $) deformation of the
amplitudes of wave packages.

The equation (\ref {eq3.35}) is supplemented by the initial condition:
\be
w (\si, {\bar \eta}, \t) _ {| \t = 0} = w _ {0} (\si, {\bar
\eta}),
\label {eq3.36}
\ee
where $ w _ {0} = 1/2\vf _ {k} + i\delta\psi _ {k}. $

The solvability of the problem (\ref {eq3.35})--(\ref {eq3.36})
in a class of functions decreasing at $ | \si | + | {\bar \eta} |
\to\infty$ can be proved similarly \cite {vak1, lk2}. This result
will be represented in the next section. It guarantees decreasing
on infinity with respect to $ \xi, {\bar\eta} $ of the right-hand
sides of the equations for $ \s {1} {v} _ {k, \o}, \ k = 1,
\ \o=\pm\o (k):$
\be
2i\o [\pt _ {\tau} + \o ^ {'} \pt _ {\xi}]
\s {1} {v} _ {k, \o} =
-6g _ {1} ^ {3,0} w _ {k, \o} | w _ {k, -\o} | ^ {2}.
\label {eq3.31}
\ee
Eliminating from the right-hand side of (\ref {eq3.23})  of the
solutions of the corresponding  homogeneous equation at $ k = 2,
\ \o= \pm\o (k)$ we shall obtain the equations for $ \s {1} {v} _ {k,
\o}: $
\bean
2i\o [\pt _ {\tau} + \o ^ {'} \pt _ {\xi}]
\s {1} {v} _ {k, \o} = 0, \quad k = 2, \ \o = \pm\o (k).
\label {eq3.32}
\eean
The equations (\ref {eq3.31}), (\ref {eq3.32}) are supplemented by initial
conditions, which are obtained from (\ref {eq3.21}), (\ref {eq3.22})
accordingly:
\be
\left [
\ba {c}
\s {1} {v} _ {k, \o (k)} \nn \\ \s {1} {v} _ {k, -\o (k)}
\ea
\right] (\xi, {\bar \eta}, \tau, \t) _ {| \tau = \t = 0} =
\left [
\ba {c}
\vf _ {+} \nn \\ \vf _ {-}
\ea
\right] (\xi, {\bar \eta}), \quad k = \pm 1,
\label {eq3.33}
\ee
\be
\left [
\ba {c}
\s {1} {v} _ {k, \o (k)} \nn \\ \s {1} {v} _ {k, -\o (k)}
\ea
\right] (\xi, {\bar \eta}, \tau, \t) _ {| \tau = \t = 0} =
\left [
\ba {c}
0 \nn \\ 0
\ea
\right], \quad k = \pm 2,
\label {eq3.34}
\ee
$$
\mbox {where} \quad\vf _ {\pm} = \mp\frac {1} {2\o (k)}
\pt _ {\tau} [\s {0} {v} _ {k, \o (k)} + \s {0} {v} _ {k, -\o (k)}]
(\xi, {\bar \eta}, \tau, \t) _ {| \tau = \t = 0}.
$$

The Cauchy problem  (\ref {eq3.32}), (\ref {eq3.34}) for $
\s {1} {v}_ {k, \o}, \ k = 2, \ \o = \pm \o (k)$ has only a
trivial solution.

The solution $ \s {1} {v} _ {k, \o} = \s {1} {v} _ {k, \o } (\si
_ {\pm}, \bar {\eta}, \tau, \t), \ k = 1, \  \o = \pm \o (k)$ of
(\ref {eq3.31}) with the initial condition (\ref {eq3.33}) is
written out in an explicit form:
$$
\s {1} {v} _ {k, \o} = \vf _ {\pm} (\si _ {\pm}, \bar {\eta}) +
2i\delta w _ {k, \o} (\si _ {\pm}, \bar {\eta}, \t)
\int\limits _ {0} ^ {\tau} |w_{k, -\o} | ^ {2} (\si _
{\pm} \pm2\o ^ {'} \mu,
\bar {\eta}, \t) d\mu,
$$
which is uniformly bounded with respect to $\tau$.

The functions $ \s {1} {v} _ {k, \o}, \ k = 1,2, \ \o = \pm\o
(k), $ are defined on this step with exactitude up to the
solutions of the corresponding homogeneous equations (\ref
{eq3.31}), (\ref {eq3.32}). Dependence of these functions from
the second slow scale $\t$ can be defined on the following steps.
But it is not necessary for formal constructions which are given
here.

After the carried out elemination of the secular terms the second
correction satisfies to the equation:
\be
\ba{ll}
[\pt _{t}^{2}-\pt _{x}^{2}+\nu \pt _{x}^{4}]\s{2}{v}=&
\sum \limits _{k=\pm 3}\sum
\limits _{\o=\pm \o (1);\pm 3\o (1)}f_{k,\o}(\xi, {\bar \eta}, \tau, \t)
e^{ikx-i\o t}+\\ & + \sum \limits _{k=\pm 1}\sum
\limits _{\o=\pm 3\o (1)}f_{k,\o}(\xi, {\bar \eta}, \tau, \t)
e^{ikx-i\o t},
\ea
\label{eq4.01}
\ee
$$
\mbox {where} \quad f _ {k, \o} =
\left\{
\ba{ll}
\mu (2-k)(w_{k,\bar{\o}})^{2}w_{-k,\bar{\o}},&
\quad k=\pm1,\ \bar {\o}=\pm\o(1), \ \o=3\bar{\o}; \\
-k\mu (w_{k/3,\bar{\o}})^{3}, &
\quad k=\pm3,\ \bar {\o}=\pm\o(1), \ \o=3\bar{\o}; \\
-k\mu (w_{k/3,\o})^{2}w_{k/3,-\o}, &
\quad k=\pm3,\ \o=\pm\o(1).
\ea
\right.
$$

The particular solution of (\ref {eq4.01}) is constructed by Fourier method:
\bea*
\s{2}{v} &=&
\sum \limits _{k=\pm3}\sum \limits _{\o=\pm \o (1);\pm 3\o (1)}
[-\o^{2}+k^{2}+\nu k^{4}]^{-1} f_{k,\o}(\xi, {\bar \eta}, \tau,
\t) e^{ikx-i\o t}+\\ &+& \sum \limits _{k=\pm1}\sum \limits
_{\o=\pm 3\o (1)} [-\o^{2}+k^{2}+\nu k^{4}]^{-1}
f_{k,\o}(\xi, {\bar \eta}, \tau, \t)e^{ikx-i\o t}.
\eea*

Thus we constructed  three terms of FAS of the Cauchy problem
(\ref {eq3.10})--(\ref {eq3.11}) at $\ve\to 0$ for long times
$0\leq t\leq {\cal O}(\ve^ {-2}), \ \forall (x, {\bar y}) \in
{\Bbb R}^{n}$. The amplitudes of the leading term of the
asymptotics are taken from (\ref {eq3.35}) and (\ref {eq3.25})
which are supplemented by the initial conditions (\ref {eq3.36})
and (\ref {eq3.26}) accordingly.

As we mentioned before, the construction of FAS can be continued
for deriving higher corrections $\s {q} {v}, \ q\geq 3 $. Here we
constructed three terms because it is sufficient for
justification of the leading term of the asymptotic expansion.

\section*{3 The investigation of standard problems}

It is known that the existence of the solution of the Cauchy
problem for the multidimensional NLS in Sobolev spaces $H^{s}
({\Bbb R}^{n})$ can be proved by different methods \cite {zhiber,
sh3}. However we can't obtain justification of the leading term
of the asymptotics in $H^{s}({\Bbb R}^{n})$. Here justification
is obtained with severe constraints, namely, with analyticity of
input data with respect to spatial variables.

The results of this section of solvability of standard problems
(\ref {eq3.35})--(\ref {eq3.36}) and (\ref {eq3.25})--(\ref
{eq3.26}) represent the generalization of the Cauchy-Kovalevskoi
theorem in Ovsyannikov's style \cite{ovs1}.

{\bf Definition 1.} Here we introduce Banach spaces $ {\cal P} _
{\beta, p} $ of functions $U(m,{\bar l})$  with the finite norm
$$
\parallel U \parallel _{p}
=\sup\limits_{m, {\bar l}} \Big[ (1+\vert m \vert +\vert  {\bar l} \vert)^{p}
e^{\beta (1+\vert m \vert+\vert  {\bar l} \vert)}\vert U(m,
{\bar l})
\vert
\Big],\quad p\geq(n+1),\ \beta>0;
$$
capital letters are used for Fourier-images:
$$
U (m, {\bar l}) = \int\limits _ {{\Bbb R} ^ {n}} u (\xi,
{\bar
\eta}) e ^ {-i (m\xi + < {\bar l} \cdot {\bar \eta} >)} d\xi d {\bar \eta}.
$$
It is necessary to notice, that similar spaces with exponential
weights in Fourier pre-images correspond to analytical functions
in layer $ S (\beta)=\left\{\vert Im\xi \vert, \vert Im\eta _ {2}
\vert, \dots, \vert Im\eta _ {n}\vert \leq\beta \right\}$ \cite {ovs1}.

The solution of the Cauchy problem (\ref {eq3.25})--(\ref
{eq3.26}) can be written out in an explicit form:
$$
v_{0}(\xi,{\bar \eta},\tau)=(2\pi)^{-n}
\int\limits_{{\Bbb R}^{n}}
\Phi_{0}(m, {\bar l})\cos (\sqrt{m^{2}+| {\bar l}|^{2}}\tau)
e^{i(m\xi+< {\bar l}\cdot{\bar \eta}>)}dmd{\bar l}.
$$
As a result we have the following
\begin {theor} \hskip-2mm {\bf}.
Let's suppose  $ \vf _ {0} (\xi,{\bar \eta}) $ is the analytical
function in layer $ S(\beta _ {0})\ (\beta_{0} = const > 0) $:
Fourier-image $
\Phi _ {0} (m, { \bar l}) \in {\cal P}_{\beta, p}. $
Then the Cauchy problem (\ref{eq3.25})--(\ref {eq3.26}) has the
analytical solution with respect to $ \xi, {\bar
\eta} $ in layer $ S (\beta _ {0}) \ (\beta _ {0} = const > 0):$
Fourier-image belongs to $ {\cal P} _ {\beta, p} $.
\label {th2}
\end {theor}
In particular, we deduce from here that the solution of the Cauchy problem
(\ref {eq3.25})--(\ref {eq3.26}) is uniformly bounded with respect to $\tau$.

Similar statement is valid for the problem (\ref {eq3.35})--(\ref {eq3.36}):
\begin {theor} \hskip-2mm {\bf}.
Let's suppose  $ \vf _ {k}, \psi _ {k} (\xi, {\bar \eta}), k =
\pm1, $ are the analytical functions in layer $ S (\beta _ {0}) \
(\beta _ {0} = const > 0) $: Fourier-images belong to  $ {\cal P}
_ {\beta, p}.$ There exist values $ \t _ {0}, \ \beta \in (0,
\beta _ {0}):$ $ \forall\ \t\in [0, \t _ {0}] $ the
Cauchy problem (\ref {eq3.35})--(\ref {eq3.36}) has a unique
solution, which is analytical with respect to  $ \xi, {\bar \eta}
$  in layer $S(\beta)$: Fourier-image belongs to ${\cal
P}_{\beta, p}.$
\label {th1}
\end {theor}
The proof is based on the Fourier transformation with respect to
spatial variables $\xi, {\bar\eta} $ and completely similar to
\cite {vak1, lk2}.

The analyticity of solution of (\ref {eq3.31}) for the amplitudes
of the first correction results from the analyticity of the
right-hand side of this equation. However we will later need
estimates of Fourier-images. Therefore the following statement
will be useful:
\newtheorem {lemma} {Lemma}
\begin {lemma}
\hskip-2mm {\bf}.
Let's suppose the function $u=u (\xi, {\bar \eta}, \tau), $ satisfies to the
equation:
$$
[\pt_{\tau}+\zeta\pt_{\xi}]u=v(\si_{+},{\bar
\eta})w(\si_{-},{\bar \eta}),
\quad\si_{\pm}=\xi\mp\zeta\tau,\quad \zeta=\hbox{const}.
$$
Then the following estimate is valid for  Fourier-image $ U = U (m, {\bar l},
\tau)$:
$$
\parallel U
\parallel _ {p} \le
M _ {0} \parallel V \parallel _ {p} \cdot\parallel W
\parallel _ {p}, \quad M _ {0} = \mbox {const} > 0, \
\forall \ V, W \in {\cal P} _ {\beta, p}.
$$
\label {l1}
\end {lemma}

{\it Proof.} Fourier-image $ U = U (m, {\bar l}, \tau) $ can be
written out in an explicit form (sign $ \star $ denotes
convolution with respect to $ (m, {\bar l})):
$
$$
U = V\star\Big (W\frac {\sin m\zeta\tau} {m\zeta} \Big) =
\int\limits _ {{\Bbb R} ^ {n}} V (m-m _ {1}, \bar {l} -\bar {l} _ {1})
W (m _ {1}, \bar {l} _ {1}) \frac {\sin m _ {1} \zeta\tau}
{ m _ {1} \zeta} dm _ {1} d\bar {l} _ {1}.
$$
Let's divide this convolution integral by sum $J _ {1}$ and $ J _
{2} $ with integration due to domains $
\Omega _ {1} =
\{| m _ {1} | > 1, \ \bar {l} _ {1} \in {\Bbb R} ^ {n-1} \}, \quad
\Omega _ {2} =
\{| m _ {1} | \leq 1, \ \bar {l} _ {1}
\in {\Bbb R} ^ {n-1} \} $
accordingly. In what follows we set for being brief:
$
\rho (m, {\bar l})
= (1 + \vert m \vert + \vert {\bar l}
\vert)^{p}e^{\beta (1 + \vert m \vert + \vert {\bar l} \vert)}.
$

Using the norm of space $ {\cal P} _ {\beta, p}, $ we estimate
integral $ J _ {1} $:
$$
| J _ {1} | \leq | \zeta | ^ {-1} \int\limits _ {{\Bbb R} ^ {n}}
| V (m-m _ {1}, \bar {l} -\bar {l} _ {1}) |
| W (m _ {1}, \bar {l} _ {1}) | dm _ {1} d\bar {l} _ {1}
\leq
$$
$$
\leq | \zeta | ^ {-1} \parallel V \parallel _ {p}
\parallel W \parallel _ {p}
\int\limits _ {{\Bbb R} ^ {n}}
\rho ^ {-1} (m-m _ {1}, \bar {l} -\bar {l} _ {1})
\rho ^ {-1} (m _ {1}, \bar {l} _ {1}) dm _ {1} d\bar {l} _ {1}
\leq
$$
$$
\leq M _ {1} | \zeta | ^ {-1} \parallel V \parallel _ {p}
\parallel W \parallel _ {p}
\rho ^ {-1} (m, \bar {l}), \quad
M _ {1} = \mbox {const} > 0.
$$
Let's estimate integral $J _ {2} $:
$$
| J _ {2} | \leq
\int\limits _ {\Omega _ {2}} | V (m-m _ {1}, \bar {l} -\bar {l} _ {1}) |
| W (m _ {1}, \bar {l} _ {1}) | | \frac {\sin m _ {1} \zeta\tau}
{ m _ {1} \zeta} | dm _ {1} d\bar {l} _ {1} \leq
$$
$$
\leq \parallel V \parallel _ {p}
\parallel W \parallel _ {p}
\int\limits _ {\Omega _ {2}}
\rho ^ {-1} (m-m _ {1}, \bar {l} -\bar {l} _ {1})
\rho ^ {-1} (m _ {1}, \bar {l} _ {1}) | \frac {\sin m _ {1} \zeta\tau}
{ m _ {1} \zeta} | dm _ {1} d\bar {l} _ {1}
\leq
$$
$$
\leq M _ {2} \parallel V \parallel _ {p}
\parallel W \parallel _ {p}
\rho ^ {-1} (m, \bar {l}) \int\limits _ {0} ^ {1}
\frac {\sin m _ {1} \zeta\tau} {m _ {1} \zeta} dm _ {1} \leq
$$
$$
\leq \pi M _ {2} | \zeta | ^ {-1} \parallel V \parallel _ {p}
\parallel W \parallel _ {p}
\rho ^ {-1} (m, \bar {l}), \quad M _ {2} =
\mbox {const} > 0.
$$
The result of the obtained estimations is the validity of lemma
\ref {l1}.

\section * {4 Justification of the asymptotics}

In this section we aim to justify an asymptotic expansion. We
must prove that the constructed segment of series (\ref {eq3.14})
gives an asymptotic expansion at $\varepsilon \to 0 $ for some
solution of the input problem (\ref {eq3.10})--(\ref {eq3.11})
for long times $0\le t\le T\varepsilon^ {-2}$.

{\it The proof of the theorem 1.} \ Let's seek for the exact
solution of the problem (\ref {eq3.10})--(\ref {eq3.11}) in the
form of:
$$
u(x, {\bar y}, t, \varepsilon) = \ve \hat {v} +
\ve ^ {2} z (x, \xi, {\bar \eta}, t, \ve),
$$
where the function $\hat {v} = \s {0} v + \ve\s {1} v + \ve ^
{2}\s {2} v $ was constructed in section 2.

Substituting this expansion into equations (\ref {eq3.12}), (\ref
{eq3.13}) we shall obtain the following Cauchy problem for the
function $ z (x, \xi, {\bar\eta}, t, \ve) $ :
\be
[ \pt _ {t} ^ {2} - (\pt _ {x} + \ve\pt _ {\xi}) ^ {2} -\ve ^ {2} \Delta _ {{\bar
\eta}} +
\nu (\pt _ {x} + \ve\pt _ {\xi}) ^ {4} + \nu \ve ^ {4} \pt _ {{\bar \eta}} ^ {4}] z =
\ve ^ {2} f,
\label {eq4.02}
\ee
\be
\left [
\ba {c}
z \nn \\ \pt _ {t} z
\ea
\right] _ {| t = 0} =
\left [
\ba {c}
z _ {1} \\ z _ {2}
\ea
\right] (x, \xi, {\bar \eta}, \ve).
\label {eq4.03}
\ee
Here
$$
f = f _ {0} + D _ {x} f _ {1} + \ve < \nabla _ {{\bar \eta}}
\cdot\bar {f} >,
\quad \bar {f} = (f _ {2}, \dots, f _ {n}),
$$
$$
f _ {0} = f _ {0} (x, \xi, {\bar \eta}, t, \ve), \quad f _ {j} =
f_ {j} (z _ {x}, z _ {\xi}, \nabla _ {{\bar
\eta}} z, x, \xi, {\bar \eta}, t, \ve), \ j = \overline {1, n};
$$
we select the known function:
$$
f_ {0} = 2 [\pt
_ {t} \pt _ {\tau} -\pt _ {x} \pt _ {\xi} + 2\nu \pt _ {x} ^ {3} \pt
_ {\xi}] \s {2} {v} + [\pt _ {\tau} ^ {2} + 2\pt _ {t} \pt
_ {\t} -\Delta _ {\xi {\bar \eta}} +
6\nu\pt _ {x} ^ {2} \pt
_ {\xi} ^ {2}] (\s {1} {v} + \ve\s {2} {v}) +
$$
$$
+ 2 [\pt
_ {\tau} \pt _ {\t} + 2\nu \pt _ {x} \pt
_ {\xi} ^ {3}] \hat {v} +
\ve [\pt _ {\t} ^ {2} + \nu \pt _ {\xi} ^ {4} +
 \pt _ {{\bar \eta}} ^ {4}] (\hat {v} -v _ {0} + \ve\s {1} {v} +
\ve ^ {2} \s {2} {v}) +
g _ {1} ^ {3,0} \pt _ {\xi} (\s {0} {v} _ {x}) ^ {3};
$$
and functions which depend on $z$:
$$
f_{1} = g _ {1} (D _ {x} [\hat {v} + \ve z], \nabla _ {{\bar
\eta}} [\hat {v} + \ve z]) -
g_{1} ^ {3,0} (\s {0} {v} _ {x}) ^ {3} + \nu\pt _ {\xi} ^ {3} v
_ {0};
$$
$$
f_{j} = g _ {j} (D _ {x} [\hat {v} + \ve z], \nabla _ {{\bar
\eta}} [\hat {v} + \ve z]) + \nu\pt _ {\eta _ {j}} ^ {3} v _ {0}; \
j = \overline {2, n}.
$$

In fact we should prove solvability and estimate of the solution
of the nonlinear problem (\ref {eq4.02})--(\ref {eq4.03}) for
residual. Note that in this problem there is present one
"superfluous" spatial variable $\xi$. The relation with the input
problem may be obtained at contraction $\xi=\ve x.$ A similar
method (extension of dimensionality of the problem) separates
periodicity with respect to $x$ from smoothness with respect to
$\xi$ in solution.

The proof of solvability of the problem (\ref
{eq4.02})--(\ref{eq4.03}) is based on the Fourier transformation
with respect to variables $x,\xi, {\bar \eta}$. Thus we pass from
the partial differential equation to the system of ordinary
differential equations for the Fourier-image of residual $Z (k,
m, {\bar l}, t,\ve):$
\be
[ \pt _ {t} ^ {2} + \lambda ^ {2}] Z =
\ve ^ {2} F,
\label {eq4.04}
\ee
with the initial conditions
\be
\left [
\ba {c}
Z \nn \\ \pt _ {t} Z
\ea
\right] _ {| t = 0} =
\left [
\ba {c}
Z _ {1} \\ Z _ {2}
\ea
\right] (k, m, {\bar l}, \ve).
\label {eq4.05}
\ee
Here $ \lambda ^ {2} = (k + \ve m) ^ {2} + \ve ^ {2} | {\bar l} |
^ {2} + \nu (k + \ve m) ^ {4} + \nu\ve ^ {4} {\bar l} ^ {4}; \
{\bar l} = (l _ {2}, \dots, l _ {n}), \ {\bar l} ^ {4} = l ^ {4}
_ {2} + \cdots + l ^ {4} _ {n}; $ capital letters are used for
Fourier-images:
$$
Z (k, m, {\bar l}) = \sum\limits _ {k\in {\Bbb Z}} \int\limits _ {{\Bbb
R} ^ {n}} z (x, \xi, {\bar \eta}) e ^ {-i (kx + m\xi + < {\bar l} \cdot {\bar
\eta} >)} d\xi d {\bar \eta}.
$$

The right-hand side $F$ in (\ref {eq4.04}) consists of three
addends:
$$
\ba {ll}
F = F _ {0} (k, m, {\bar l}, t, \ve) & + i (k + \ve m) F _ {1}
(kZ, mZ, {\bar l} Z, k, m, {\bar l}, t, \ve) + \\ & + i\ve <
{\bar l} \cdot {\bar F} (kZ, mZ, { \bar l} Z, k, m, {\bar l}, t,
\ve)
>,
\ea
$$
and contains various convolutions of functions $ kZ, mZ, {\bar l}
Z $ and some known ones. In what follows sign $\star$ denotes
convolution with respect to $(k, m, {\bar l}).$

Solution $Z_ {0} = Z _ {0} (k, m, {\bar l}, t,\ve)$ corresponding
to the known right-hand side $F_{0}$ can be written out in an
explicit form:
\be
[ \pt _ {t} ^ {2} + \lambda ^ {2}] Z _ {0} = \ve ^ {2} F _ {0}.
\label {eqF0}
\ee
We choose the solution of (\ref {eqF0}) which satisfies to the
following initial conditions:
\be
\left [
\ba {c}
Z _ {0} \nn \\ \pt _ {t} Z _ {0}
\ea
\right] _ {| t = 0} =
\left [
\ba {c}
Z _ {1} \\ Z _ {2}
\ea
\right] (k, m, {\bar l}, \ve).
\label {eqF0in}
\ee
Let's seek for the solution of the problem (\ref {eq4.04})--(\ref
{eq4.05}) in the form of: $ Z = H + Z _ {0}.$ Then we shall
obtain the following Cauchy problem for $H$:
\be
[ \pt _ {t} ^ {2} + \lambda ^ {2}] H = i (k + \ve m) F _ {1} + i\ve < {\bar l}
\cdot {\bar F} >,
\label {eq4.06}
\ee
\be
\left [
\ba {c}
H \nn \\ \pt _ {t} H
\ea
\right] _ {| t = 0} =
\left [
\ba {c}
0 \\ 0
\ea
\right].
\label {eq4.07}
\ee

The linear term with the factor $\lambda^{2}$ in the left-hand
side of (\ref{eq4.06}) may be excluded by  replacement of the
function using fundamental solution of the corresponding
homogeneous one, namely:
$$
H (k, m, {\bar l}, t, \ve) = Ae ^ {-i\lambda t} + Be ^ {i\lambda t},
$$
where $ A, B = A, B (k, m, {\bar l}, t, \ve). $
$$
\mbox{If}\quad
{ \bf R} = \left[
\ba {c}
A \\ B
\ea \right], \quad
{ \bf G} =
\left [
\ba {c} G ^ {+} \\ G ^ {-}
\ea \right], \ \
\hbox {where} \ \ G ^ {\pm} = \mp \frac {\Big ((k + \ve m)
F _ {1} + \ve < {\bar l} \cdot {\bar F} > \Big)} {2\lambda} e ^ {\pm i\lambda
t},
$$
we obtain the following differential equation for vector function
$ {\bf R}$:
\be
{ \bf R} _ {t} = \ve ^ {2} {\bf G}.
\label {eq4.08}
\ee
The initial condition for (\ref {eq4.08}):
\be
{ \bf R} \vert _ {t = 0} = 0.
\label {eq4.09}
\ee
After all the problem (\ref {eq4.08})--(\ref {eq4.09}) is reduced
to the integral equation for $ {\bf R} (k, m, {\bar l}, t,
\varepsilon) $:
\be
{ \bf R} (k, m, {\bar l}, t, \varepsilon) =
\varepsilon ^ {2} \int\limits ^ {t} _ {0}
{ \bf G} (k, m, {\bar l}, \mu, {\bf R} (k, m, {\bar l}, \mu,
\varepsilon), \varepsilon) d\mu.
\label {eq4.10}
\ee

Here we use a scale of Banach spaces for the proof of the
existence of solution, analogously \cite {ovs1, vak1, lk2,
lksh2}. Their elements are the functions with the  band of
analyticity  $ \beta (t) > 0 $. This band is narrowed down in due
course. The corresponding norms are defined through the
Fourier-images with exponential weights.

{ \bf Definition 2.} Here we introduce Banach spaces $ {\cal H} _
{\beta, p} $ of functions $ U (k, m, {\bar l}) $ with the finite
norm
$$
\parallel U \parallel _ {\beta, p}
 = \sup\limits _ {k, m, {\bar l}} \Big [(1 + \vert k \vert +
\vert m \vert + \vert {\bar l} \vert) ^ {p} e ^ {\beta (1 + \vert k \vert +
\vert m \vert + \vert {\bar l} \vert)} \vert U (k, m, {\bar l}) \vert \Big],
\quad  p\geq (n + 2), \ \beta > 0.
$$

\begin {lemma} \hskip-2mm {\bf}.
The convolution operator is a bounded one in $ {\cal H} _ {\beta,
p}:$
$$
\parallel U \star V
\parallel _ {\beta, p} \le
M _ {0} \parallel U \parallel _ {\beta, p} \cdot\parallel
V\parallel _ {\beta, p}, \quad M _ {0} = \mbox {const} > 0, \quad
\forall\quad U, V \in {\cal H} _ {\beta, p}.
$$
\end {lemma}

\begin {lemma} \hskip-2mm {\bf}.
The convolution operator is a Lipschitzian one in $ {\cal H}
_ {\beta, p}:\  \forall M _ {1} < \infty $
$$
\parallel U \star U - V \star
V \parallel _ {\beta, p}
\leq 2 M _ {0} M _ {1}
\parallel U - V \parallel _ {\beta, p},
$$
$$
\forall \quad U, V:\quad \parallel U
\parallel _ {\beta, p}, \parallel V \parallel _ {\beta, p}
\leq M _ {1}.
$$
\end {lemma}
The proof of these statements may be obtained analogously \cite
{vak1, lk2}.

The results of it are the boundedness and Lipschitzian of any
degrees of convolution in $ {\cal H} _ {\beta, p}$. We shall
define a degree of convolution by relation $ (\star U) ^ {1} = U,
\quad (\star U)
^ {j} = (\star U) ^ {j-1}
\star U\ (j = 2,3, \dots). $

\begin {lemma} \hskip-2mm {\bf}.
The degree of  convolution is a bounded and Lipschitzian operator
in $ {\cal H}
_ {\beta, p}:\ \forall M _ {1} < \infty, \
\forall j = 1,2, \dots $
$$
\| (\star U) ^ {j} \| _ {\beta, p} \leq M _ {0} ^ {j-1}
\| U \| _ {\beta, p};
$$
$$
\| (\star U) ^ {j} - (\star V) ^ {j} \| _ {\beta, p}
\leq j (M _ {0} M _ {1})^{j-1} \| U-V \| _ {\beta, p};
$$
$$
\forall \quad U, V:\quad \parallel U
\parallel _ {\beta, p}, \parallel V \parallel _ {\beta, p}
\leq M _ {1}, \quad M _ {0} = \mbox {const} > 0.
$$
\end {lemma}
The proof is given by induction with respect to $j$.

We introduce the space $ C ([0, T\ve ^ {-2}] \times [0, \ve _
{0}]; {\cal H}_{\beta, p}) $ of functions $ U (t, \ve) $ which
are continuous with respect to $ (t, \ve) \in [0, T\ve ^ {-2}]
\times [0, \ve _ {0}] $ with the finite norm:
$$
\parallel U \parallel = \sup \limits _ {t\in [0, T\ve ^ {-2}]}
\sup \limits _ {\ve\in [0, \ve _ {0}]}
\parallel U \parallel _ {\beta, p}.
$$

\begin {lemma} \hskip-2mm {\bf}.
The solution $ Z _ {0} (k, m, {\bar l}, t, \ve) $ of the Cauchy
problem (\ref {eqF0})--(\ref {eqF0in})  belongs to $ C ([0, T\ve
^ {-2}] \times [0, \ve _ {0}]; {\cal H} _ {\beta, p})$ provided
theorems \ref {th2} and \ref {th1}.
\label {l7}
\end {lemma}

{\it Proof.} We can write out the solution of ordinary
differential equations (\ref {eqF0})  with initial conditions
(\ref {eqF0in}) in an explicit form, but here it is not
necessary. It is worth mentioning, that the right-hand side
$F_{0}$ of this equation contains Fourier-images constructed
above FAS, hence, $ F _ {0} \in C ([0, T\ve ^ {-2}] \times [0,
\ve _ {0}]; {\cal H} _ {\beta, p}). $ As all addends in $ F _ {0}
$ have the form $ \tilde {z} (k, m, {\bar L},\ve) \exp (i\o t) $
with $\o\not = 0 $, where $ \tilde {z} (k, m, {\bar L},\ve)
\in C ([0, T\ve ^ {-2}] \times [0, \ve _ {0}]; {\cal H} _ {\beta, p}), $
the solution $ Z _ {0} (k, m, {\bar l}, t, \ve) \in C ([0, T\ve
^ {-2}]\times [0, \ve _ {0}]; {\cal H} _ {\beta, p}). $
Lemma \ref {l7} is proved.

In what follows we introduce Banach spaces $ {\cal {\bf H}} _
{\beta, p} $ of vector functions ${\bf R}(k,m,$ ${\bar l},t,\ve)$
with continuous components with respect to $ k, m, $$ {\bar l},
t, \ve $ and with exponential decreasing at infinity with respect
to $k,m,{\bar l} $. The norm in space $ {\cal {\bf H}} _ {\beta,
p} $ is defined as the sum of norms of the components of vector $
{\bf R} $ in space $ {\cal H} _ {\beta, p} $:
$$
\parallel {\bf R} \parallel _ {\beta, p} = \parallel
A\parallel _ {\beta, p} +
\parallel B\parallel _ {\beta, p}.
$$
Vector functions $ {\bf R} (k, m, {\bar l}, t, \ve) \in {\cal
{\bf H}} _ {\beta, p} $, which are continuously depending with
respect to $ t \in [0, T\ve ^ {-2}],\ve\in [0, \ve _ {0}] $ are
considered in Banach space $ {\bf C}_ {p} = C ([0, T\ve ^ {-2}]
\times [0, \ve _ {0}]; {\cal {\bf H}}_ {\beta, p}) $ with the
norm:
$$
\parallel {\bf R} \parallel _ {
{ \bf C} _ {p}} = \sup _ {t, \ve} \parallel {\bf R} (k, t, m, {\bar l}, \ve)
\parallel _ {\beta, p}.
$$
\noindent
In what follows the index $ \beta = \beta (\t) $ will depend on $
\t =
\varepsilon ^ {2} t $.

The considered equations reduce to integration operator from
(\ref {eq4.10}), which contains both convolution degrees and
exterior factors $ (k + \ve m) / (2\lambda) $; $\ve
l_{j}/(2\lambda)$, which are uniformly bounded (the majorant is
1/2).

Factors $ k $, $ m $, $ l _ {j} $ coming from operators $ \pt _
{x} $,
$
\pt _ {\xi}, $ $ \pt _ {\eta _ {j}} $ are unbounded ones in
$ {\cal {\bf H}} _ {\beta, p} $. But they will be bounded from $
{\cal {\bf H}} _ {\beta, p} $ in $ {\cal {\bf H}} _ {\beta, p-1}
$. On the other hand, the integration operator
$$
I [{\bf R}] = \ve ^ {2} \int\limits ^ {t} _ {0} {\bf R} (k, m, {\bar
L}, \mu, \ve) d\mu
$$
is linear bounded from $ {\bf C} _ {p-1} $ in $ {\bf C} _ {p} $ with $ \beta =
\beta _ {0} -\beta _ {1} \ve ^ {2} t \ (\beta _ {0}, $$ \beta _ {1} = \hbox
{const} > 0) $ on the functions continuously depending with
respect to $t,\ve$. This integration operator has estimate
analogously \cite {vak1, lk2}:
$$
\parallel I [{\bf R}] \parallel _ {
{ \bf C} _ {p}} \le \beta _ {1} ^ {-1} \parallel {\bf R}
\parallel _ {{\bf C} _ {p-1}},
\quad
\forall \quad t \le
\beta _ {0} \beta _ {1} ^ {-1} \ve ^ {-2}.
$$
If now we choose a sufficiently large constant $ \beta _ {1} $,
the composition $I [(k + m + l _ {j}) {\bf R}] $  operator of
multiplication with respect to $ (k + m + l _ {j}) $ with the
integrated one will be the contractive operator in scale of
spaces $ {\cal {\bf H}}_ {\beta, p} $ on functions, continuously
depending with respect to $ t, \ve $. A similar situation is in
case with general operator from (\ref {eq4.10}). Integrals in the
right-hand side (\ref {eq4.10}) will be bounded and Lipschitzian
operators in $ {\bf C}
_ {p} $. Therefore the integration operator from (\ref {eq4.10}) is
contractive in $ {\bf C} _ {p} $ with $ \beta = \beta
_ {0}
-\beta _ {1} \ve ^ {2} t \ (\beta _ {0}, $$ \beta _ {1} = \hbox {const} > 0)
$. This is what ensures a local solvability of integral equations
(\ref {eq4.10}) in $ {\bf C}_{p} $. Thus the upper bound of
existence interval $T\le\beta _ {0} \beta _ {1}^ {-1} \ve ^ {-2}
$  is defined from the condition $
\beta = \beta _ {0}
-\beta _ {1} \ve ^ {2} t \geq 0\ (\beta _ {0}, \beta _ {1} = \hbox {const} >
0) $. Thus the leading term of the asymptotics in (\ref {eq3.14})
is justified. Theorem \ref{th0} is proved.

The author is grateful to L.A.Kalyakin for useful discussions.

\end{document}